\documentclass{appolx}
\usepackage{epsfig}

\begin{document}

\title{Ground-state phase diagram of geometrically frustrated Ising-Heisenberg
       model on doubly decorated planar lattices%
\thanks{Presented at CSMAG'07 Conference, Ko\v{s}ice, 9-12 July 2007}%
}
\author{L. \v{C}anov\'a, J. Stre\v{c}ka, J. Dely and M. Ja\v{s}\v{c}ur
\address{Department of Theoretical Physics and Astrophysics, Faculty of
Science, P.~J.~\v{S}af\'arik University, Park Angelinum 9, 040 01
Ko\v{s}ice, Slovak Republic}}

\maketitle

\begin{abstract}
Ground-state phase diagram of the mixed spin-$1/2$ and spin-$1$
Ising-Heisenberg model on doubly decorated planar lattices is
examined using the generalized decoration-iteration transformation.
The main attention is devoted to the comparison of the ground-state
properties of the quantum Ising-Heisenberg model and its
semi-classical Ising analogue.
\newline
\end{abstract}

\section{Introduction}
Frustrated quantum models have attracted a great research interest
during the last three decades especially due to their extraordinary
diverse ground-state behaviour, which often arises as a result of
mutual interplay between the geometric frustration and quantum
fluctuations~\cite{Die94}. Despite a considerable effort, the
geometric frustration and its effect on magnetic properties of
frustrated spin systems have not been fully elucidated, yet. Owing
to this fact, exactly solvable geometrically frustrated models can
serve as useful models for in-depth understanding of this
phenomenon~\cite{Str04}. In this work, we shall provide the exact
solution for the special class of geometrically frustrated
Ising-Heisenberg planar models, which could be potentially helpful
in the examination of the geometric frustration.

\section{Model system and its solution}
\begin{figure}[htp]
     \begin{center}
     \vspace{-2.5cm}
       \includegraphics[angle = 0.0, width = 0.75\textwidth]{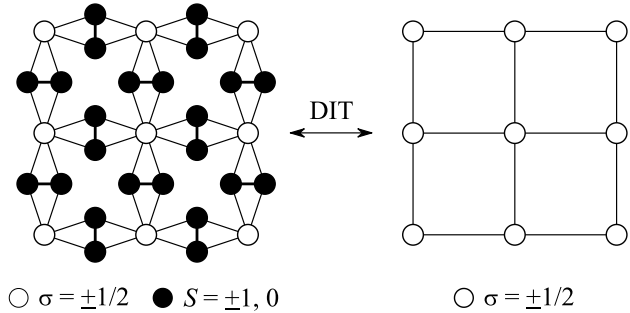}
       \vspace{-0.75cm}
       \caption{\small Diagrammatic illustration of the used mapping transformation.
       The mixed spin-$1/2$ and spin-$1$ Ising-Heisenberg model on the doubly
       decorated square lattice is mapped on the simple spin-$1/2$ Ising model on
       the corresponding lattice. Empty (full) circles denote lattice positions of
       the Ising (Heisenberg) spins.}
       \label{fig1}
     \end{center}
   \end{figure}
Consider the mixed spin-$1/2$ and spin-$1$ Ising-Heisenberg model on
doubly decorated planar lattices as schematically depicted on the
left-hand side of Fig.~\ref{fig1} for the case of doubly decorated
square lattice. The total Hamiltonian of the model under
investigation can be written in the form $\hat{\cal{H}}_{\rm IH} =
\sum_{k = 1}^{Nq/2}\hat{\cal{H}}_k$, where the summation is carried
out over all bonds of the original (undecorated) lattice of $N$
atoms with the coordination number $q$. The bond Hamiltonian
$\hat{\cal{H}}_k$ is given by the formula
\begin{eqnarray}
\hat{{\cal{H}}}_k &=& J_{\rm H}[\Delta(\hat{S}_{k1}^{x}
\hat{S}_{k2}^{x} + \hat{S}_{k1}^{y} \hat{S}_{k2}^{y}) +
\hat{S}_{k1}^{z}\hat{S}_{k2}^{z}] - D[(\hat{S}_{k1}^{z})^2 +
(\hat{S}_{k2}^{z})^2] \nonumber\\&+& J_{\rm I}(\hat{S}_{k1}^{z} +
\hat{S}_{k2}^{z})(\hat{\sigma}_{k1}^{z} +\hat{\sigma}_{k2}^{z}),
\label{eq:Hk}
\end{eqnarray}
where $\hat{\sigma}_{k}^{z}$ and $\hat{S}_{k}^{\alpha}$ ($\alpha =
x, y, z$) denote spatial components of the spin-$1/2$ and spin-$1$
operators, $J_{\rm H}$, $J_{\rm I}$ stand for the exchange
interactions between the nearest-neighbour Heisenberg spins, the
Heisenberg and Ising spins, respectively, $\Delta$ marks the XXZ
exchange anisotropy and $D$ denotes the single-ion anisotropy. Using
the commutation rule between different bond Hamiltonians and
applying the generalized decoration-iteration mapping
transformation~\cite{Syo72}, one obtains the equality
\begin{eqnarray}
{\cal{Z}}_{\rm IH}(\beta, J_{\rm H}, J_{\rm I}, \Delta, D) =
A^{Nq/2}{\cal{Z}}_{\rm I}(\beta, R),
\label{eq:Z}
\end{eqnarray}
which establishes an exact mapping relationship between the
partition function ${\cal{Z}}_{\rm IH}$ of the Ising-Heisenberg
model and the partition function ${\cal{Z}}_{\rm I}$ of the
spin-$1/2$ Ising model on the corresponding lattice (see
Fig.~\ref{fig1}). Notice that the mapping relation (\ref{eq:Z}) is
universal and valid regardless of the lattice topology or space
dimensionality of the investigated system. Besides, it also allows
direct calculation of all relevant physical quantities, which are
useful for the understanding of the magnetic behaviour in the whole
parameter space. In this regard, exact results for the
Ising-Heisenberg model on several decorated planar lattices can be
obtained, because of known exact solutions for partition functions
of many spin-$1/2$ Ising planar lattices~\cite{Syo72}.

\section{Results and concluding remarks}
Now, let us take a closer look at the ground-state behaviour of the
mixed-spin Ising-Heisenberg model with the antiferromagnetic (AF)
exchange parameters $J_{\rm I}$ and $J_{\rm H}$ ($J_{\rm I}>0$,
$J_{\rm H}>0$). Note that this sign choice closely relates to our
endeavour to match the situation in the geometrically frustrated
Ising-Heisenberg planar models. Before discussing the results in
detail, it is important to remark that the ground-state behaviour
discussed below is rather general due to the universality of
Eq.~(\ref{eq:Z}), since this mapping relation holds regardless of
the lattice coordination number $q$.

\begin{figure}[htp]
\begin{center}
\vspace{-2.5cm}
\includegraphics[angle = 0.0, width = 0.95\textwidth]{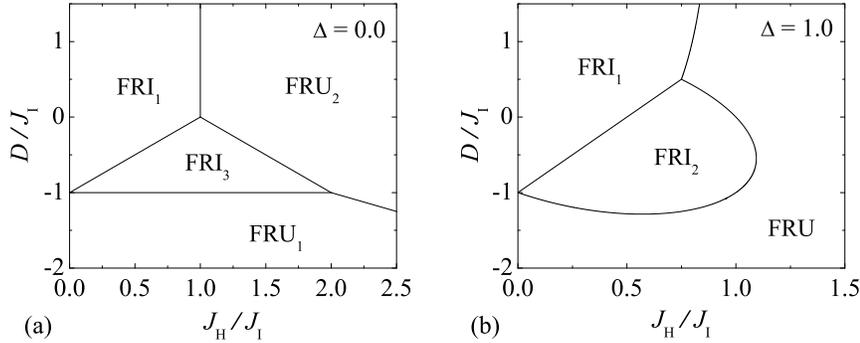}
\vspace{-1.5cm}
\caption{\small Ground-state phase diagrams of the
         spin-$1/2$ and spin-$1$ Ising-Heisenberg planar model in the
         $J_{\rm H} - D$ space for $\Delta = 0.0$ and $\Delta = 1.0$.}
\label{fig2}
\end{center}
\end{figure}

For illustration, two ground-state phase diagrams are displayed in
the $J_{\rm H}-D$ space in Fig.~\ref{fig2} for $\Delta = 0.0$ and
$\Delta = 1.0$. As one can see, the mutual competition between the
parameters $J_{\rm H}$, $J_{\rm I}$, $D$ and $\Delta$ gives rise to
six possible ground states. Spin order to emerge within each sector
of the phase diagrams can be unambiguously characterized by the
wavefunctions:
\begin{eqnarray}
|{\rm FRI}_1\rangle \!\!&=& \!\! \prod_{i=1}^{N} |-\rangle_{i}\!\!
\prod_{k=1}^{Nq/2}\!\! |1, 1\rangle_{k}; \,\,\,\,\, |{\rm
FRI}_2\rangle = \!\prod_{i=1}^{N} |-\rangle_{i}\!\!
\prod_{k=1}^{Nq/2}\!\! \frac{\,1}{\sqrt{2}} \Bigl(|1,0\rangle_{k} -
|0,1\rangle_{k}\!
   \Bigr);
\label{eq:FRI2}
\\
|{\rm FRI}_3\rangle \!\! &=& \!\! \prod_{i=1}^{N} |-\rangle_{i}\!\!
\prod_{k=1}^{Nq/2}\!\! |1, 0\rangle_{k}
\,\,\,\,\,\,\,\textrm{or}\,\,\,\,\,\, |{\rm FRI}_3\rangle =
\prod_{i=1}^{N} |-\rangle_{i}\!\! \prod_{k=1}^{Nq/2}\!\!
|0,1\rangle_{k}; \label{eq:FRI3}
\\
|{\rm FRU}\rangle  \!\! &=&  \!\! \prod_{i=1}^{N}
|\pm\rangle_{i}\!\! \prod_{k=1}^{Nq/2}\!\! \frac{\,1}{\,2} \Bigl[\,
a_{+} \Bigl( |1,-1\rangle_{k} + |-\!1,1\rangle_{k}\!\Bigr) -
\sqrt{2}\,a_{-} |0,0\rangle_{k}\Bigr], \label{eq:FRU}
\end{eqnarray}
where $a_{\pm} = \sqrt{1 \pm \delta^{-1}}$, $\delta^2 = 1 + 8(J_{\rm
H}\Delta)^2\!/(J_{\rm H}+2D)^2$. In above, the first product is
taken over all Ising spins ($|\pm \rangle$ stands for $\sigma^z =
\pm 1/2$) and the second one runs over all Heisenberg spin pairs
($|\pm\! 1,\! 0 \rangle$ is assigned to $S^z = \pm 1, 0$). It is
quite evident from the aforelisted eigenfunctions that ${\rm
FRI}_{1}$ represents the semi-classically ordered ferrimagnetic
phase with the antiparallel alignment between the nearest-neighbour
Ising and Heisenberg spins. Besides, another two ferrimagnetic
phases ${\rm FRI}_{2}$ and ${\rm FRI}_{3}$ can be found in the
quantum Ising-Heisenberg model and in its semi-classical Ising
version, respectively. Both these phases appear as results of the
competition between the AF exchange interaction $J_{\rm H}$ and the
easy-plane single-ion anisotropy ($D<0$) and the only difference
between them lies in the quantum entanglement that emerges just
within ${\rm FRI}_{2}$ (${\rm FRI}_{3}$ is classical ferrimagnetic
phase without entanglement). Finally, several frustrated phases can
be found in the ground state depending on whether the exchange
anisotropy is zero or non-zero.
If $\Delta \neq 0.0$, then the unique frustrated phase ${\rm FRU}$
can be detected as shown in Fig.~\ref{fig2}(b) for the particular
case $\Delta = 1.0$. In this phase, the Heisenberg spin pairs
exhibit quantum entanglement of three spin states $|0,0\rangle$ and
$|\pm1,\mp1\rangle$, whereas their probability amplitudes depend on
a mutual ratio between $J_{\rm H}$, $\Delta$ and $D$ according to
Eq.~(\ref{eq:FRU}). If $\Delta = 0.0$, the ground-state phase
diagram consists of two different frustrated phases ${\rm FRU}_{1}$
and ${\rm FRU}_{2}$ instead of unique ${\rm FRU}$ one. In the former
frustrated phase ${\rm FRU}_{1}$, all Heisenberg spin pairs reside
the 'non-magnetic' $|0,0\rangle$ spin state, while in the latter
phase ${\rm FRU}_{2}$, they reside one of two intrinsic AF spin
states. Note that both the frustrated phases ${\rm FRU}_{1}$ and
${\rm FRU}_{2}$ can be regarded as limiting cases of the frustrated
phase ${\rm FRU}$.

In conclusion, the investigated Ising-Heisenberg model exhibits an
interesting ground-state behaviour including the peculiar geometric
spin frustration. The more detailed examination concerned with the
finite-temperature phase diagram of this spin system will be
presented in the near future.
\\ \\*
{\bf Acknowledgments:} This work was supported under the scientific
grants VEGA~1/2009/05, LPP-0107-06 and VVGS~PF~02/2007/F.

\end{document}